\documentclass{elsart}

\usepackage{graphicx}
\usepackage{amssymb}
\usepackage{wasysym}

\newcommand{\be}{\begin{equation}}
\newcommand{\ee}{\end{equation}}
\newcommand{\bea}{\begin{eqnarray}}
\newcommand{\eea}{\end{eqnarray}}

\newcommand{\sav}{s_{\hbox{\tiny av}}}
\newcommand{\sigmaav}{\sigma_{\hbox{\tiny av}}}
\newcommand{\tn}{\tilde{n}}
\newcommand{\Ndis}{N_{\hbox{\tiny dis}}}
\newcommand{\Ragg}{\tilde{R}_{\hbox{\tiny agg}}}
\newcommand{\Gunc}{G_{\hbox{\tiny unc}}}

\begin{document}
\begin{frontmatter}

\title{Fluctuations and scaling in models for particle aggregation}

\author{Daniele Vilone}
\address{Dipartimento di Fisica, Universit\`a di
Roma ``La Sapienza'' and Center for Statistical Mechanics and Complexity,
INFM Unit\`a Roma 1, P.le A. Moro 2, 00185 Roma, Italy}
\author{Claudio Castellano$^*$}
\address{Dipartimento di Fisica, Universit\`a di
Roma ``La Sapienza'' and Center for Statistical Mechanics and Complexity,
INFM Unit\`a Roma 1, P.le A. Moro 2, 00185 Roma, Italy}
\address{Istituto dei Sistemi Complessi, CNR,
Via dei Taurini 9, 00185 Roma, Italy}

\corauth{Corresponding author. {\it E-mail address:}
claudio.castellano@roma1.infn.it}
\author{Paolo Politi}
\address{Istituto dei Sistemi Complessi, Consiglio Nazionale delle
Ricerche, Via Madonna del Piano 10, 50019 Sesto Fiorentino, Italy}

\date{\today}

\begin{abstract}
We consider two sequential models of deposition and aggregation for
particles. The first model (No Diffusion) 
simulates surface diffusion through a deterministic capture
area, while the second (Sequential Diffusion) 
allows the atoms to diffuse up to $\ell$ steps.
Therefore the second model incorporates more fluctuations than the
first, but still less than usual (Full Diffusion) 
models of deposition and diffusion
on a crystal surface. We study the time dependence of the
average densities of atoms and islands and the island size distribution.
The Sequential Diffusion model displays a nontrivial steady-state
regime where the island density increases and the island size
distribution obeys scaling, much in the same way as the standard
Full Diffusion model for epitaxial growth.
Our results also allow to gain insight into the role of different types
of fluctuations.
\end{abstract}

\end{frontmatter}

\section{Introduction}

The diverse morphologies that occur as result of the growth
of epitaxial thin films are not only of great interest
from the point of view of technological applications, but also
a fascinating subject for basic science.
How the interplay between simple microscopic processes (deposition,
diffusion, aggregation) gives rise to complex patterns on
larger scale is a nontrivial question that has attracted a lot
of interest, in particular since the spectacular refinement of
experimental techniques has provided an unprecedented control
over real growth processes~\cite{Books}.

The growth of the first monolayer has a special importance:
islands form a template over which further growth proceeds, and its
morphology may have a strong influence on the nucleation and growth
of subsequent layers.
For this reason submonolayer growth has been intensely
investigated~\cite{reviews}
and much progress has been done in the comprehension
of the different stages in the growth process.

From the theoretical point of view, the preferred approach has been
the use of rate equations~\cite{Venables84,Stoyanov81}.
They allow to understand well the behavior of mean quantities, like the
scaling of island density with respect to time, flux intensity and
temperature~\cite{Bartelt92,Bales94}.
However the application of rate equations to characterize more in detail
the morphology of the growing layer, by computing the Island Size Distribution
(ISD), has turned out to be much more difficult, even in the simplest case
of irreversible growth.

Kinetic Monte Carlo (KMC) numerical simulations support the conclusion
that, instead of depending separately on the coverage $\theta$ and the ratio
between the diffusion constant of adatoms $D$ and the incoming flux $F$,
the ISD collapses, in the steady-state precoalescence regime,
on a unique function when
sizes are scaled by their mean value $\sav(\theta,D/F)$.
The naive application of rate equations with constant capture
numbers~\cite{Bartelt92,Bales94} yields unsatisfactory results:
the ISD scales with $\sav$, but with a
scaling function that diverges for a finite value of its
argument and is zero beyond it, at odds with the smooth
function found in KMC simulations.
The origin of the failure has been traced to the mean-field assumption
that the typical environment of an island is independent of its size.
Instead, larger islands tend to have larger capture zones and
as a consequence, larger capture numbers.
This has led to very sophisticated improvements of the original theory
in order to overcome this difficulty and provide a correct description
of the evolution of the system.
Bartelt and Evans~\cite{Bartelt96} have related the form of the
scaling function for the ISD to a generic scaling form for the
capture numbers. This last quantity is then
determined via the solution of an evolution equation for the
joint probability distribution of island sizes and capture zones,
under plausible assumptions and
simplifications~\cite{Mulheran00,Evans01,Evans02,Mulheran04}.
Other approaches include a general analysis in the limit of infinitely
fast adatom diffusion~\cite{Vvedensky00} and a self-consistent
coupling of rate equations for island size with equations for capture
zones~\cite{Amar01}.

Very recently some doubts have been cast over this conceptual
framework and in particular about the very existence of scaling
for the ISD in the steady-state.
Authors of Ref.~\cite{Ratsch05} have claimed that scaling of the ISD with
respect to coverage does not rigorously occur for point island models, while
it does for extended island models. The ultimate reason for this
difference is ascribed to the fact that for point islands nucleation
continues throughout the steady-state, while it is almost completely
suppressed for extended islands. According to Ref.~\cite{Ratsch05}
continuous nucleation would lead asymptotically to a configuration
where the size of the capture zones is delta-like distributed and this
would imply the absence of scaling.

In any case, it is evident that considerable progress has been achieved
via the refinement of the rate equation approach.
In the present work, we take a different path.
Instead of trying to refine further the application of the rate equation
approach to the standard model for submonolayer growth, we investigate,
by means of suitably modified models, what is the effect of the
different types of fluctuations that enter submonolayer growth.
Work along this line has been previously performed in Ref.~\cite{Ratsch00},
where it is shown that the ISD obtained in KMC simulations is
well reproduced by a deterministic continuum model, the level-set
method~\cite{Gyure98}, with the sole additional
ingredient that the spatial positioning of new islands occurs stochastically
with probability depending on the local value of the adatom diffusion field.

In our paper, we study two simplified atomistic models that switch
off some sources of fluctuations: this allows, on the one hand, to
investigate the precise role of each of them, and on the other hand,
to understand the origin of the scaling property of the ISD.
Both models are sequential, in the sense that the deposition of an adatom
takes place only when the one deposited before does not move any more.
In the first case [No Diffusion model (ND)], a deposited adatom remains
where it is deposited, unless it is within a fixed distance from an island
that incorporates it.
In the Sequential Diffusion (SD) model, an adatom performs a random walk
immediately after deposition. If it is not incorporated after a fixed
number of steps it stops forever, becoming the seed of a new island.
Both models have already been used in the context of multilayer
growth~\cite{Biehl98,Ahr00} and the first (No Diffusion) model
has been recently studied for submonolayer growth in 
one-dimension~\cite{Politi05}.

The two models are in some sense successive approximations of the standard
(point-island) Full Diffusion (FD) model~\cite{Bartelt92} that is thought
to include all relevant physical ingredients for irreversible
submonolayer growth in the precoalescence regime.
In such a model several mechanisms give rise to spatial and temporal
fluctuations: the deposition flux, the diffusion of adatoms, the nucleation
of new islands.
The two models considered in this paper shut off some of these noise sources.
Temporal fluctuations in the flux are disregarded because both models are
sequential.
The only source of fluctuations in the ND model is provided by the spatial
fluctuations of the flux; all the rest of the dynamics is deterministic.
The SD model has an additional type of spatial noise, associated to the
diffusive motion of adatoms.
The main difference with respect to the standard model is in the nucleation
process, that requires in practice only one particle instead of two.

In the following we study the temporal evolution of mean
adatom and island densities and the scaling properties of the ISD.
It turns out that the behavior of the No Diffusion model is far,
even from the qualitative point of view, from the FD model.
Its steady-state is characterized by a complete lack of nucleation
events and, as a consequence, by a ``trivial'' scaling of the ISD.
This is due to the fully deterministic treatment of adatom
diffusion.
The Sequential Diffusion model, instead, captures most of the
physics of FD model, exhibiting a steady-state with
continuous nucleation and a nontrivial scaling of the ISD.
The comparison between the two models makes apparent that
scaling of the ISD can have two distinct origins: either the
complete suppression of island nucleation in the steady-state 
or the persistence of nucleation events nontrivially coupled
to the evolution of capture numbers. The shape of the ISD scaling
function mirrors the different mechanisms at work.

\section{The Models}

We consider two different models for irreversible submonolayer growth,
characterized by an increasing amount of fluctuations. They are
point-island models, i.e. the spatial extent of islands is always
one lattice unit, island size being simply a counter attached to
them~\cite{Bartelt92}.
If $N_p$ particles have been deposited on a square substrate of
$L\times L$ sites,
the ratio $\theta=N_p/L^2$ defines the coverage $\theta$.
Point-island models are expected to describe well the behavior
of extended island models, provided the total coverage is small,
$\theta \ll 1$, so that islands do not coalesce.

We call the first model ``No Diffusion (ND) model", defined as follows.
At each time step we randomly choose a lattice site and check whether some
atoms or islands exist within a distance $\ell$.
If they do, the newly deposited adatom sticks to the closest atom or island,
otherwise it sticks permanently to the deposition site.
Since adatoms do not diffuse, the distinction between adatoms and islands
is only formal in this model:
adatoms are in all respects islands of size one.
We keep the distinction for comparison with the standard Full Diffusion model.
Random deposition is the only source of noise in the ND model.

\begin{figure}
\includegraphics[angle=0,width=13cm,clip]{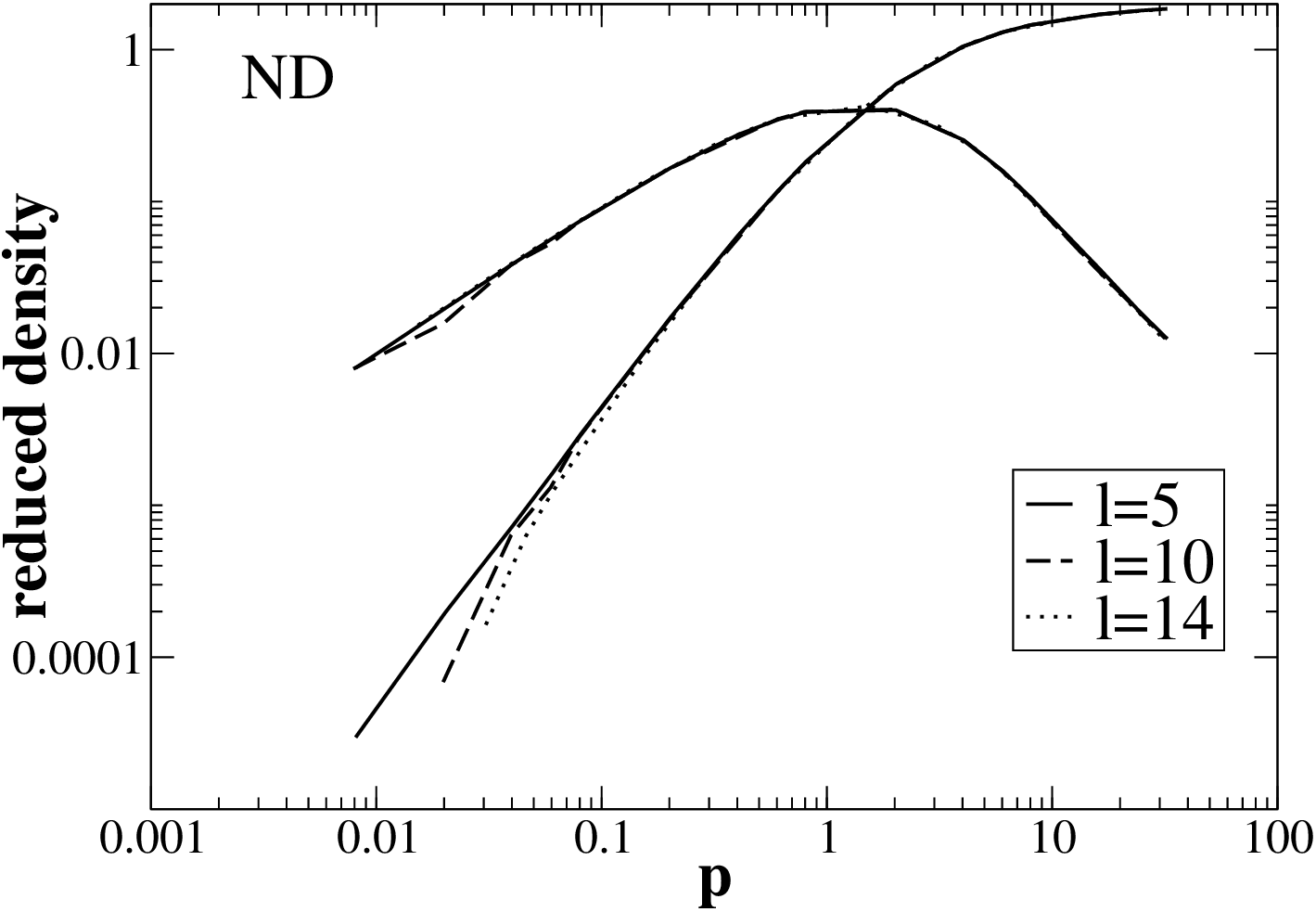}
\caption{Reduced density of adatoms $\tn_1$ (upper lines for small $p$)
and islands $\tn_{is}$ (lower lines for small $p$)
for the No Diffusion model with
$\ell=5$ (solid), $\ell=10$ (dashed), $\ell=14$ (dotted),
showing that only $p$ matters. Data are for a system of size $L=100$.}
\label{verif_p_ND}
\end{figure}

The length $\ell$ is the only length scale in the problem (apart
from the lattice size). Hence all physical quantities are the same
provided all lengths are rescaled by $\ell$ (as long as
$\ell \gg 1$, so that discreteness of the lattice can be neglected).
Fundamental quantities are the densities, defined as $n_x = N_x/L^2$,
where  $N_x$ is the number of objects of type $x$ (adatoms, islands) 
present in the system.
The properly rescaled densities are reduced densities, defined
as $\tn_x = \pi N_x/(L/\ell)^2 = c n_x$,
where $c = \pi \ell^2$ is the capture area surrounding a particle.
The physical meaning of the reduced density $\tn_x$ is
the number of particles of type $x$ per capture area.
Also the coverage depends on a length ($L$). The properly rescaled
coverage is $p=\pi N_p/(L/\ell)^2=c \theta$, i.e. the average number
of particles deposited per capture area.
The fact that the evolution of the system depends on $\theta$ and $\ell$
only through the parameter $p=c\theta$ is verified numerically for the
reduced densities in Fig.~\ref{verif_p_ND}.
As discussed in Sec.~\ref{Sec4.b.1}, for some properties this model is
equivalent to the Random Sequential Adsorption model~\cite{Evans93}.

The second model considered is the Sequential Deposition (SD) model.
We choose at random a lattice site and put there a new adatom.
This adatom diffuses randomly for at most $\ell$ steps.
If it happens to go on an occupied site, it sticks to the preexisting
island (or adatom).
If the adatom does not meet any island or adatom during its whole
random walk, it stops forever on the last site becoming an ``island''
of size 1. Also in this case we will continue to call atoms adatoms, but 
once again the distinction between such adatoms and islands is formal.

Also in the SD model all quantities do not depend separately on $\theta$
and $\ell$, rather on the combination $p=c(\ell) \theta$.
The definition of the capture area $c(\ell)$ is in this case slightly
different from the previous one: it has to be intended as a probabilistic
concept, the region surrounding an island such that a deposited
adatom is likely to be incorporated by such an island, before
ending its walk.
Hence the capture area is given by the
number $\Ndis(\ell)$ of distinct sites visited by a walker during
$\ell$ steps: $c=\Ndis(\ell)$.
In two dimensions it is well known~\cite{Hughesbook} that,
for asymptotically large $\ell$
\be
\Ndis \simeq \frac{\pi \ell}{\log(\ell)}.
\label{Ndis}
\ee

\begin{figure}
\includegraphics[angle=0,width=13cm,clip]{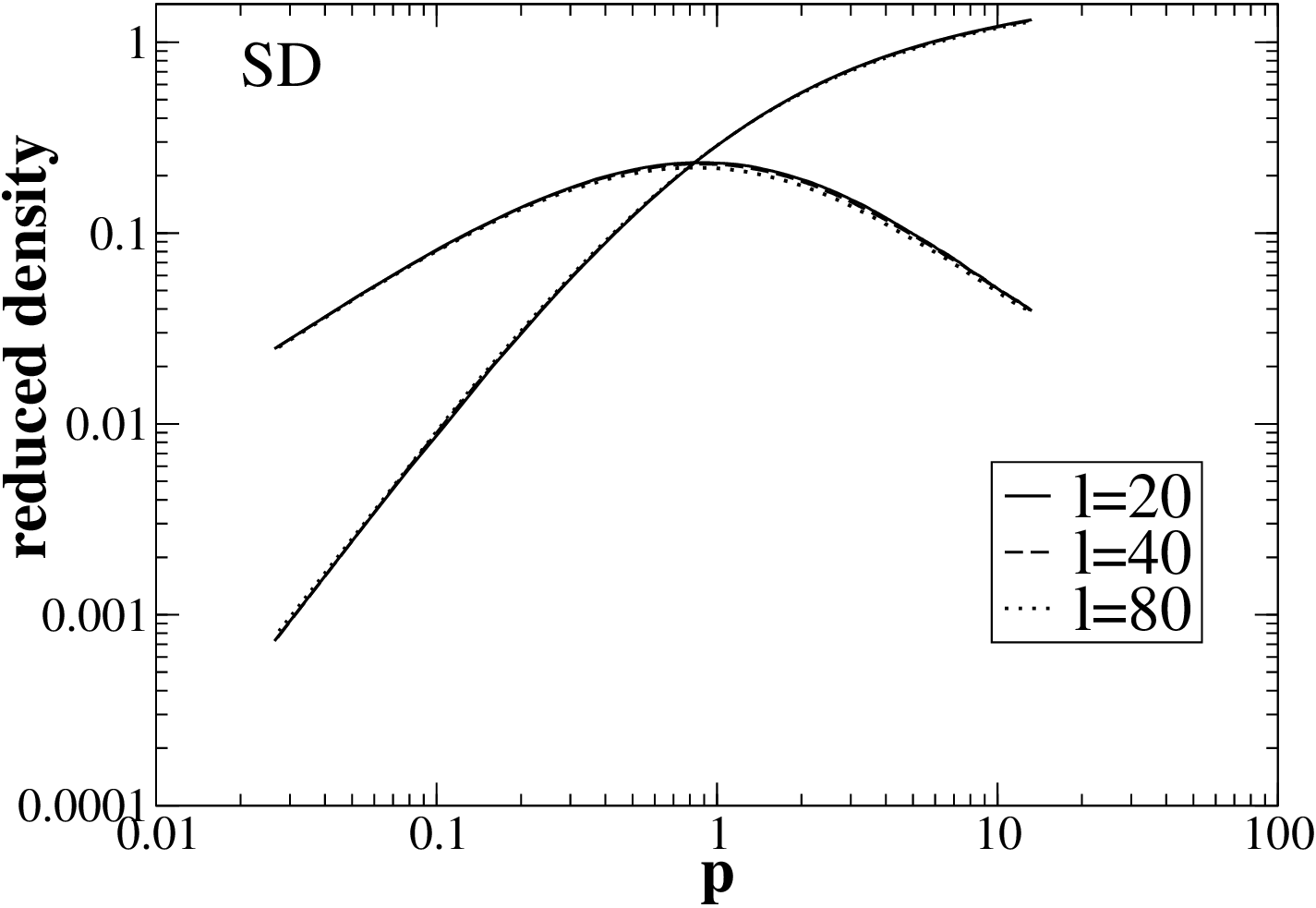}
\caption{Reduced density of adatoms $\tn_1$ (upper lines for small $p$)
and islands $\tn_{is}$ (lower lines for small $p$)
for the Sequential Diffusion model with
$\ell=20$ (solid), $\ell=40$ (dashed), $\ell=80$ (dotted),
showing that only $p$ matters. Data are for a system of size $L=200$.
The value of $p$ is computed using Eq.~(\ref{Ndis}).}
\label{verif_p_SD}
\end{figure}

The numerical proof that $p$ is the only relevant parameter for the
SD model is provided in Fig.~\ref{verif_p_SD}.
The sources of noise in this model are the random deposition and the
diffusion process of adatoms.

The Sequential Diffusion model bears some resemblance to another model
used for describing epitaxial growth in the presence
of impurities or surfactants, the spontaneous nucleation (SN) model, also
denoted by the size $i=0$ of the critical
nucleus~\cite{Chambliss94,Amar95}.
The prescription for the SN model is that at each
diffusive step adatoms have a finite probability to stop, nucleating
in this way a new island. Intuition suggests that if the probability to
stop is such that the mean number of steps taken by the adatom is $\ell$, the
model with $i=0$ might behave in a fashion very similar to the SD model.
We will discuss below whether this idea turns out to be correct.

Since in both the ND and the SD models the only parameter is $p=c \theta$,
it is important to remark that large $p$ does not imply necessarily
large coverage $\theta$.
With sufficiently large $\ell$, the asymptotic stage of these
models (see below) can occur even for $\theta \ll 1$,
i.e. when the assumption of point islands is a reasonable approximation
for extended islands.

The SD model incorporates more fluctuations than the ND model,
but it still lacks two (related) features of the models usually studied in
KMC simulations: the fact that adatoms are continuously diffusing and the
parallel, not sequential, character of the deposition process. 
If we take both into accounts, we have the standard
model for irreversible growth for point islands~\cite{Bartelt92},
that we call Full Diffusion (FD) model.
A flux $F$ of adatoms is deposited at random on the sites of a lattice.
Once deposited, each adatom diffuses (with diffusion constant $D$)
until it meets a preexisting island (thus being incorporated)
or another adatom (nucleating a new island of size 2).
This model has been thoroughly studied~\cite{Bartelt92}.
We will not simulate it but we will often refer to its well known behavior.
It is not sequential and it contains all sources
of noise: temporal and spatial randomness in the deposition process plus
the whole fluctuations of adatom diffusion and island nucleation.

\section{Behavior of average densities}

\begin{figure}
\includegraphics[angle=0,width=13cm,clip]{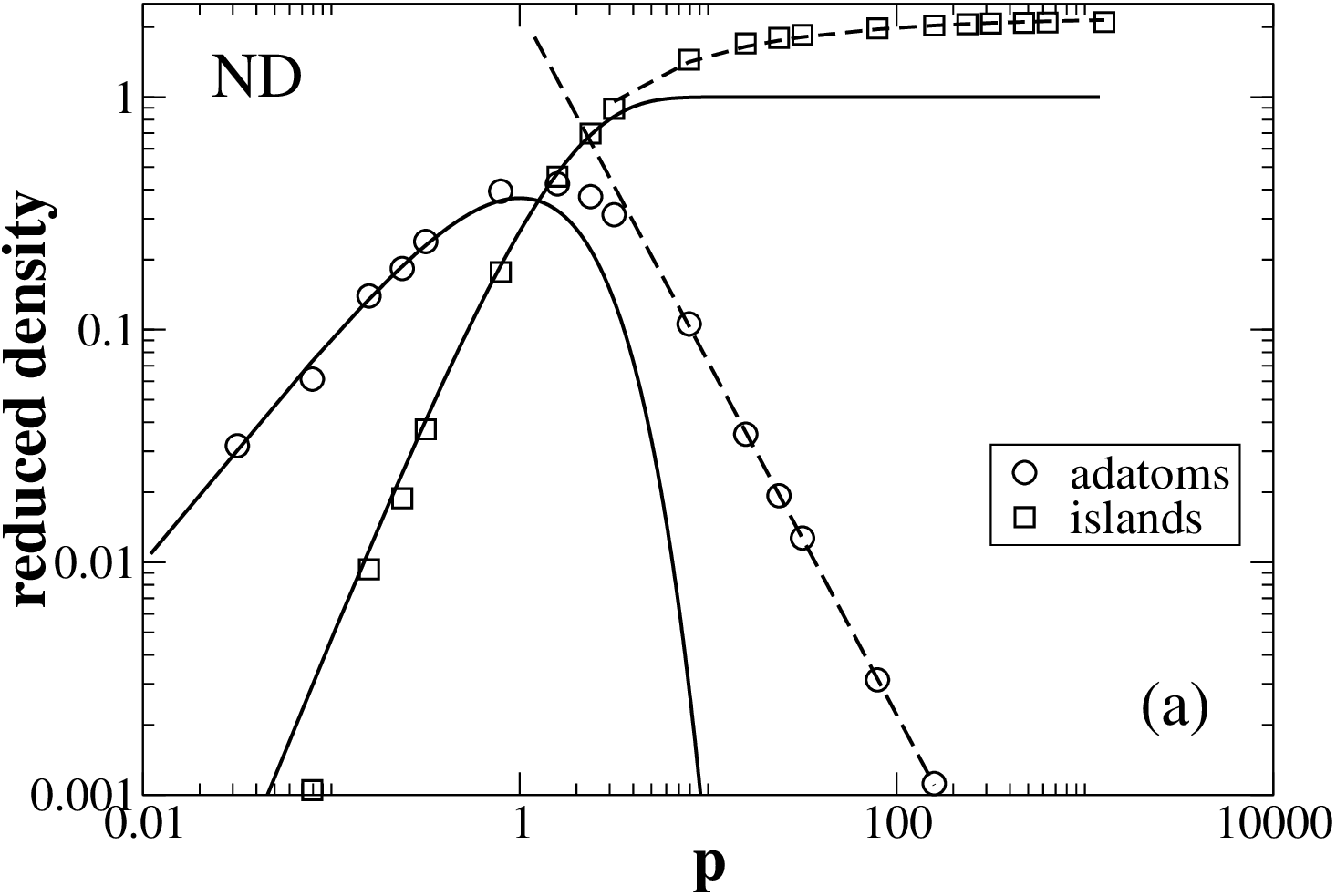}
\includegraphics[angle=0,width=13cm,clip]{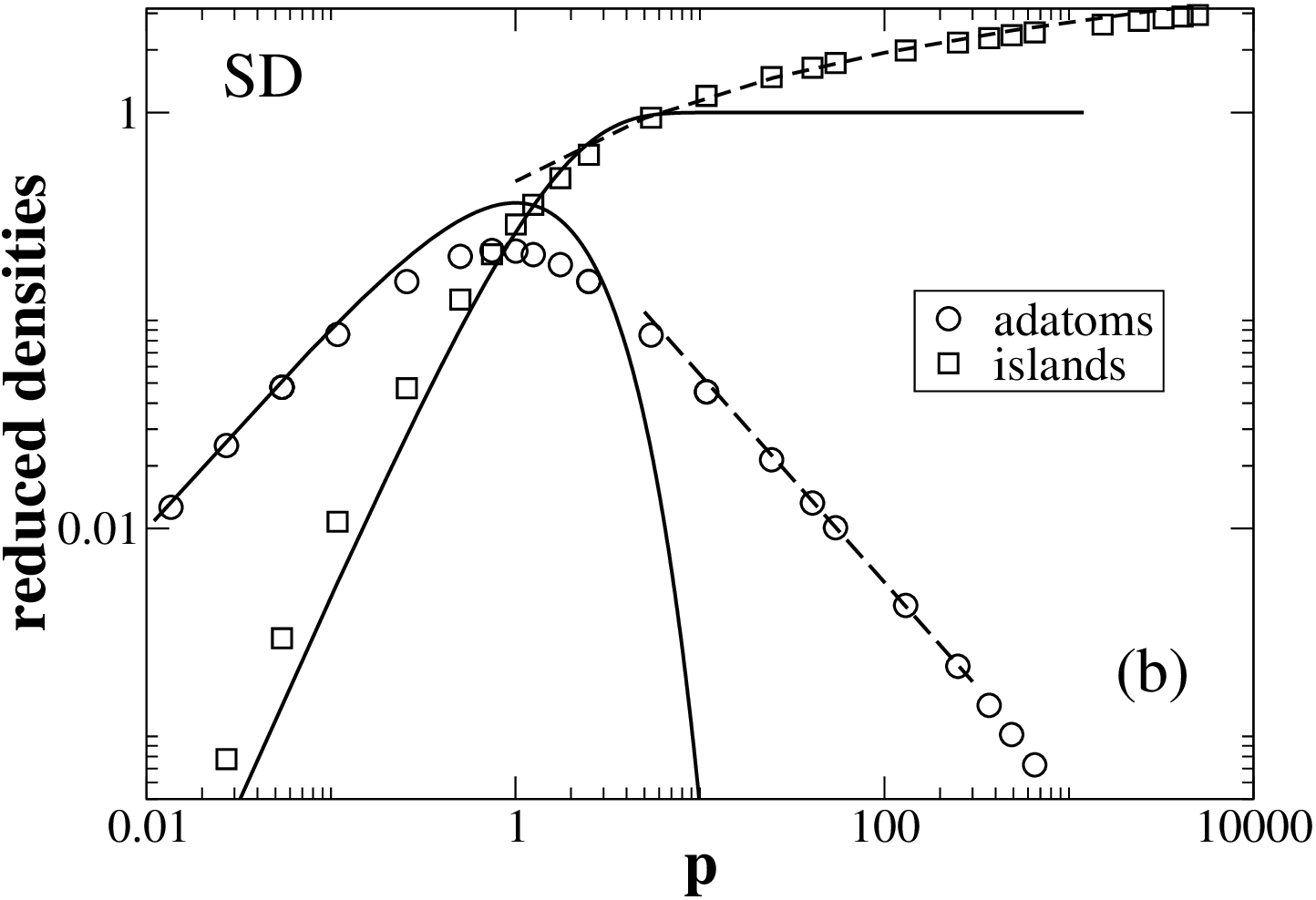}
\caption{Reduced density of adatoms $\tn_1$ and islands
$\tn_{is}$ for the ND model (a) and the SD model (b).
Solid lines are the solutions~(\protect\ref{meanf_sol}) of the rate
equations. The decreasing dashed lines are power-law decays as $p^{-3/2}$ (a)
and $p^{-1}$ (b). 
The increasing dashed lines are: (a) a fit to $\tilde n_{is}=
1/\tilde A_\infty - c_1/\sqrt{p}$, with $\tilde A_\infty\approx 0.45$
and $c_1\approx 2.2$ (see Sec.~\ref{Sec4.b.1});
(b) the analytic curve $\tilde n_{is}=(1/\alpha)\log (1+\alpha p)$,
with $\alpha\simeq 2.93$ (see Sec.~\ref{Sec4.b.2}).
Data are for $\ell=10$ (a) and $\ell=100$ (b).}
\label{dens}
\end{figure}

The simplest quantities that describe the process of submonolayer
growth are the reduced density of adatoms $\tn_1$ and
of islands $\tn_{is}$.
In Figure~\ref{dens} we report the behavior of the
average adatom and island densities as a function of the
rescaled temporal variable $p$.

In both cases a crossover at $p \approx 1$ separates two different
regimes.
For small $p$ nucleation dominates, the density of atoms
increases linearly, and the density of islands increases
quadratically, as predicted by rate equations (see next section).
At large $p$ the island density cannot decrease, because
coalescence of islands is not possible in point-island models.
For the ND model, $\tn_{is}$ reaches asymptotically a saturation value
$\tilde{n}_{is}^{\infty}$.
Instead for the SD model the island density grows
logarithmically before attaining a constant value when the
discreteness of the lattice comes into play~\footnote{The 
transition to the final asymptotic
stage for the SD model
with constant island density occurs only for values of $p \gg c$.
In such a limit, corresponding to the condition $\theta \gg 1$ for
which point island models become unphysical, 
the adatom and island densities approach
exponentially 0 and a value of the order of 1, respectively.
In our simulations we have never reached such extremely high values of $p$.}.
The density of adatoms decreases for large $p$ as $p^{-3/2}$
for the ND model and as $p^{-1}$ for the SD case.

These patterns of behavior can be qualitatively understood as follows.
At the beginning, new atoms are typically deposited in empty regions
so that their density grows linearly with the coverage.
Only occasionally a new particle lands in the capture area of a preexisting
adatom, giving rise to a new island.
The average distance among atoms or islands is very large.
With increasing $p$, atoms and island densities increase and when
$p$ becomes larger than 1 their capture areas start to overlap.
From then on the growth processes of distinct islands are no more independent
and the difference in the models results in a different evolution.
At large $p$, almost all neighbouring islands have overlapping
capture areas but they do not cover the whole lattice.
There are still some regions such that all their sites are outside the
capture area of any particle, so that the new adatoms deposited there
do not get incorporated.
Densities of atoms and islands change because these active regions are
filled by adatoms and adatoms are then transformed into islands.
While for the ND model capture areas are deterministic, so that once they
cover the whole lattice no more nucleations are possible, for the SD
model it is always possible that an adatom deposited close to an island
does not in fact meet the island during its diffusive motion.
Such adatom will stop at the end of its walk and give rise to a new island.
This difference explains why the island density tends to a constant for the
ND model, while it keeps growing (though slowly) in the SD case.
In the latter case the growth of $\tn_{is}$ ends only when the island density
is so high (of the order of 1) that new particles can only
be incorporated in existing islands.

These results must be compared to the behavior of the analogous quantities
for the FD model. When adatom diffusion is fully operative,
the density of islands grows as $\theta^{3}$ during the initial regime,
while  the adatom density increases linearly.
In the ensuing steady-state regime instead,
$n_{is}$ grows as $(F\theta/D)^{1/3}$.
Simultaneously the density of adatoms decreases as
$(F \theta D^2)^{-2/3}$~\cite{Bartelt92}.
From this comparison it turns out that, from the quantitative point of
view, both models with reduced fluctuations fail to reproduce, at least
for what concerns the exponents, the phenomenology of the FD model.
For example, the temporal exponent for the growth of $n_{is}$ during the
nucleation stage is 2, instead of 3.
This is a direct consequence of the fact that in the FD model
two mobile adatoms must meet to nucleate an island, while in the present
models nucleation involves a single adatom, that, if not immediately
incorporated after deposition, becomes immobile, in this way seeding
a new island.
However, from a more qualitative point of view, the SD model performs
rather well, since it exhibits regimes which
have the same nature of the more realistic FD model.
In particular, a steady-state regime during which islands
are nucleated is nontrivial.
This means that, after a sufficiently long time interval,
most islands have been nucleated during this regime and the
nucleations occurred during the initial stage are negligible.
This is what happens also in the FD model and will be shown to have
profound consequences for the scaling of the Island Size Distribution.

In the stationary-state the behavior of the density of islands is then
the direct consequence of the amount of spatial fluctuations present
in the nucleation process. The failure of the ND model in reproducing
such a regime indicates that the existence of an area around particles
where new nucleation events are deterministically forbidden is a
too drastic simplification.

In the two models considered here, the quantity $\ell$ determines
the value of the densities in the steady-state. Hence it plays
a role analogous to $D/F$ in the FD model.
Another difference deserves to be remarked. In the Full Diffusion model
the morphology of the system depends on two parameters, $\theta$ and $D/F$,
that can be changed independently. In the ND and SD models instead,
$\theta$ and $\ell$ are combined in the only relevant parameter $p$.
This means that the scaling property of the ISD (see below) holds
with respect to both parameters or to none.

\section{Rate equations}

\subsection{Nucleation stage $p \ll 1$}
As discussed above, for both the ND and the SD model
particles are effectively ``non-interacting'' in the limit of small $p$,
because their mean distance is so large that typically
capture areas do not overlap.
In this limit it is easy to write down the following mean-field
rate equations:
\bea
\frac{d\tilde{n}_1}{dp} & = & 1-(\tilde{n}_{is}+2\tilde{n}_1) 
\label{meanf} \\ \nonumber
\frac{d\tilde{n}_{is}}{dp} &= & \tilde{n}_1 
\eea
with initial conditions $\tn_1(0)=\tn_{is}(0)=0$.
Notice that the rate equations are the same for the two models,
the only difference being the definition of $p$.
The fact that the rate of island nucleation is proportional
to $\tn_1$ reflects the fact that only one adatom is necessary for it.

The solutions are
\bea
\tilde{n}_1(p) & = & p\exp(-p) 
\label{meanf_sol} \\
\tilde{n}_{is} & = & 1-(p+1)\exp(-p).
\nonumber
\eea

These expressions, that make sense for any $p$, are reported
in Fig.~\ref{dens} as full lines and compared with numerical results (symbols).
The comparison is very good up to $p\approx 1$, i.e. during the initial
nucleation stage.
For larger values the overlap of capture areas becomes relevant and the
above approximation is no longer valid, as expected.
The equations for the evolution of the density of islands of any size
$s\geq 2$ are
\be
\frac{d\tilde{n}_s}{dp}=\tilde{n}_{s-1}-\tilde{n}_s.
\ee
They can be solved recursively,
showing that at small coverage the size distribution is Poissonian
\be
\tn_s = {p^s \over s!} \exp(-p).
\label{n_s}
\ee
It is worth noting that the above rate equations, and therefore
their solutions, do not depend on the space dimension.
Changing the dimension simply changes the definition of the capture area $c$.

\subsection{Steady-state $p \gg 1$}

For large values of $p$ the form of rate equations must be modified
to take into account the overlap of capture areas.
Rate equations for a generic model of irreversible growth are
\bea
\frac{d\tilde{n}_1}{dp} & = & 1-2 \Ragg(1)-\sum_{s>1} \Ragg(s)
\label{Rate_generic} \\ \nonumber
\frac{d\tilde{n}_s}{dp} & = & \Ragg(s-1)-\Ragg(s)~~~~~~~~~~~~~~~~s>1
\eea
where $\Ragg(s)$ is the reduced aggregation rate for islands of size $s$.
By summing over $s$, the equation for the total density of islands 
$\tn_{is}$ is obtained
\be
\frac{d\tilde{n}_s}{dp}=\Ragg(1).
\ee
The specification of the aggregation rates $\Ragg(s)$ fully determines
the set of equations. In analogy with the procedure for
the Full Diffusion model~\cite{Bartelt96} we define the capture
numbers $\sigma_s$ by the relation
\be
\Ragg(s)= \sigma_s \tn_s.
\ee

When all capture numbers are equal to one, we recover
Eq.~(\ref{meanf}) valid during the noninteracting nucleation stage.
In Section~\ref{Sec5}, devoted to the study of the Island Size Distribution,
we will discuss the appropriate expressions for capture numbers $\sigma_s$
in the regime for large $p$ and make use of Eq.~(\ref{Rate_generic}).
Here we study the asymptotic behavior of the average densities by
means of approaches tailored on the specific model.

\subsubsection{No Diffusion model}
\label{Sec4.b.1}

When $p$ is large, almost all lattice sites are at a distance
$d<\ell$ from the closest island, i.e. they are part of a capture area.
However, there are some domains whose sites are at $d>\ell$ from every
island or adatom: we denote as $I_0$ the number of such active regions.
Similarly, we call $I_1$ the number of regions corresponding to the
capture zones of adatoms.
Finally, let us call $A_{\infty}$ the average size of
capture zones for $p\rightarrow \infty$,
and $\tilde{A}_\infty = A_{\infty}/c$ the corresponding reduced quantity.
 
Let now $N_p$ be the number of particles deposited and $\Delta y(\theta)$
the mean size of the active regions.
At very high $p$ we can assume that the deposition of an adatom in
an active region (occurring with probability $I_0 \Delta y(\theta)/L^2$)
results in the complete cancellation of such active region, so that
$I_0$ is reduced by one.
If instead an adatom is deposited in a region close to a preexisting
adatom (and this happens with probability $A_{\infty} I_1/L^2$)
a new island will be nucleated and $I_1 \to I_1-1$.
Hence the following equations hold
\bea
\frac{dI_0}{dN_p} & = & - {\Delta y I_0 \over L^2} 
\label{highp} \\ \nonumber
\frac{dI_1}{dN_p} & = & {\Delta y I_0 \over L^2} - 
{A_{\infty} \over L^2} I_1.
\eea
 
\begin{figure}
\includegraphics[angle=0,width=13cm,clip]{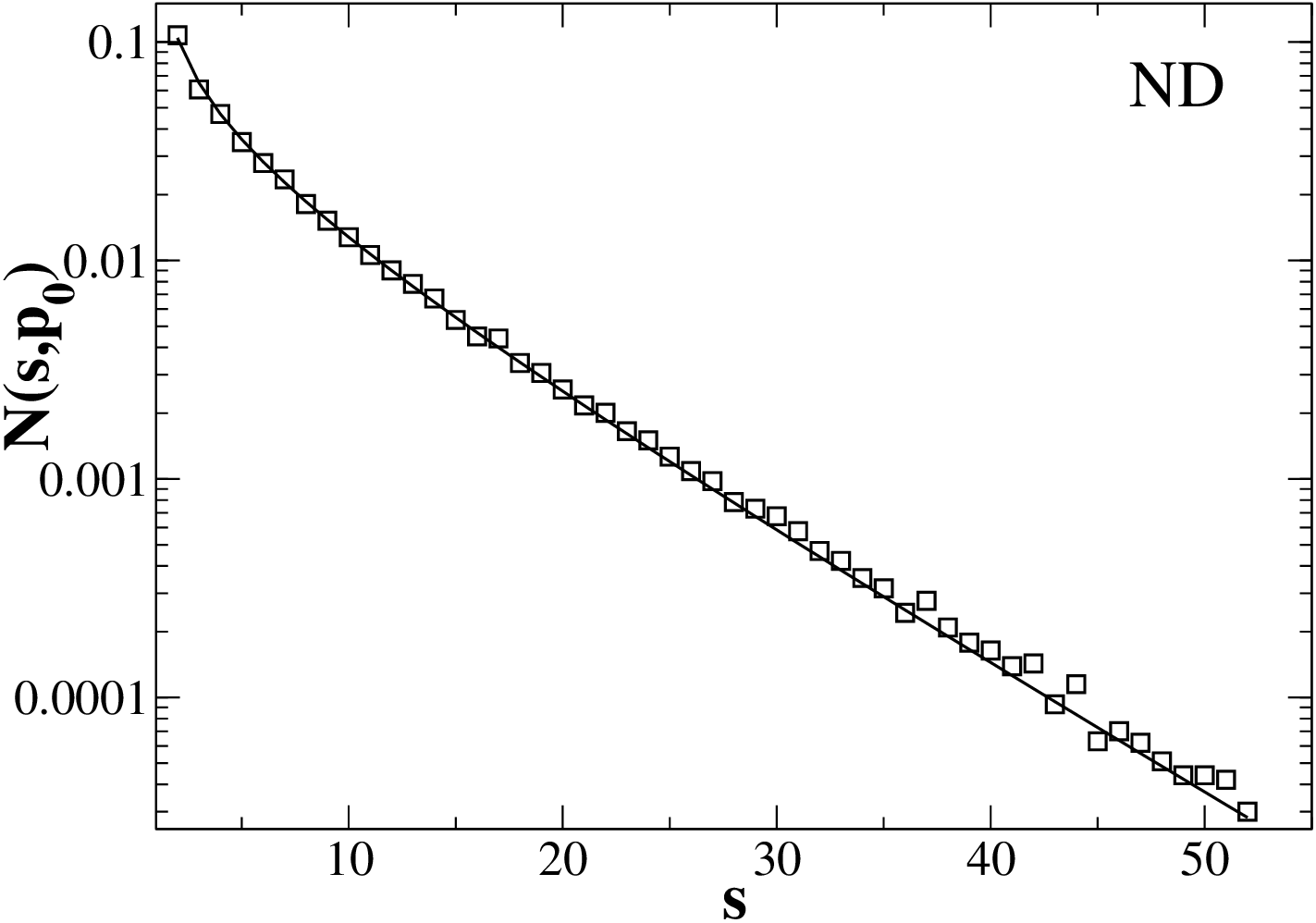}
\caption{
Distribution of the size of active regions at $p\simeq 245$.
Symbols: numerical results for the ND model with $\ell=10$.
Line: analytical results according to Eq.~(\ref{distr}).}
\label{fig_Ps}
\end{figure}

The crucial point is the determination of the form of 
$\Delta y$ for $p \gg 1$.
For large $p$ active areas can be considered to be independent.
The evolution equation for the number $N(s,p)$ of active
areas of size $s$ is $dN(s,p)/dp = -s N(s,p)/c$ so that
\be
N(s,p) = N(s,p_0) \exp[-s(p-p_0)/c]
\ee
where $p_0$ is the value of $p$ (of the order of 1) after which 
active areas can be considered as independent.
The form of $N(s,p_0)$ depends on the dynamics during the nucleation
stage for small $p$.
For its determination it is useful to take advantage of the similarity
between the No Diffusion model and the so called
Random Sequential Adsorption (RSA) model, that is well studied in
the literature~\cite{Evans93}.
Our model can be mapped onto RSA for disks of radius $r=\ell/2$.
There are some differences in the temporal properties of the two models,
but for what concerns spatial properties the mapping is exact.
We can use the result for the size distribution~\cite{Swendsen81}
\be
N(s,p_0) \propto (s-1)^{-1/2}
\label{distr}
\ee
(see Fig.~\ref{fig_Ps} for a numerical check), so that
\be
\Delta y=
\langle s\rangle=\frac{\int_{1}^{\infty}s N(s,p)ds}
{\int_{1}^{\infty} N(s,p)ds} = 1 + {1 \over 2 \theta}.
\label{delt0}
\ee

The first term appearing in the r.h.s. of Eq.~(\ref{delt0}) reflects the
discreteness of the lattice, the obvious fact that no region
can be smaller than 1.
When active regions are on average much larger than 1 site,
the first term can be neglected.
This occurs for $\theta \ll 1$ (even if $p \gg 1$, because $\ell$ is large).
Inserting $\Delta y \approx 1/(2\theta)$ into Eqs.~(\ref{highp}) yields

\bea
\frac{dI_0}{dp}&=&-\frac{I_0}{2p} \\ \nonumber
\frac{dI_1}{dp}&=&\frac{I_0}{2p}-\tilde{A_\infty} I_1.
\label{i_eq}
\eea
As expected everything depends only on $p=c(\ell)\theta$. The solutions are 
 \bea
I_0(p)&=&\frac{c_1}{\sqrt{p}} \label{i_sol} \\
I_1(p)&=&\frac{c_1}{2\tilde{d}}\sum_{k=1}^{\infty}
\left[ \frac{(2k-1)!!}{(2\tilde{A_\infty})^{k-1}}
\frac{1}{p^{k+\frac{1}{2}}}\right]
\nonumber
\eea

If instead $\theta \gg 1$, $\Delta y(\theta) \approx 1$:
all remaining active regions have very small size
and lattice discreteness dominates.
One finds that $n_1$ decays exponentially to zero and in the same
way $n_{is}$ goes to its asymptotic finite limit.
In this limit observables depend separately on $\ell$ and $\theta$.

For $p\rightarrow \infty$ Equations~(\ref{i_sol}) imply
\be
\tilde{n}_1(p)\propto I_1\approx\frac{c_1}{2\tilde{A_\infty}}
\frac{1}{p^{3/2}}
\label{i1_approx-sol}
\ee
The power-law decay with exponent $3/2$ for the adatom density is
in agreement with simulations (see Fig.~\ref{dens}).
Since for $p \to \infty$ both $I_0$ and $I_1$ vanish, only
islands remain in the system and their capture zones cover the
whole lattice. The asymptotic value of their density is then
$\tn_{is} = 1/\tilde{A}_\infty$ with corrections, for finite $p$,
proportional to the largest of the two vanishing contributions,
$I_0 \sim p^{-1/2}$.

\subsubsection{Sequential Diffusion model}
\label{Sec4.b.2}

In the Sequential Diffusion model, a freshly deposited adatom wanders
until it sticks to a preexisting adatom, to an island or it stops
after $\ell$ random-walk steps. 
Let us denote as $\rho_1$, $\rho_{is}$ and $\rho_0$ the normalized
probabilities for these events to occur, respectively.
The evolution of the average densities for the
Sequential Diffusion model can be written as

\bea
\frac{d\tilde{n}_1}{dp}& =& \rho_0-\rho_1 \\ \nonumber
\frac{d\tilde{n}_{is}}{dp}&=&\rho_1.
\label{meanf_isl}
\eea

Notice that since preexisting adatoms and islands are perfectly equivalent
for the walker, the ratio $\rho_1/\rho_{is}$ must be equal
to the ratio of the densities $\tn_1/\tn_{is}$, so that
\bea
\rho_1& =&(1-\rho_0)\frac{\tilde{n}_1}{\tilde{n}_1+\tilde{n}_{is}} \\ \nonumber
\rho_{is}& =&(1-\rho_0)\frac{\tilde{n}_{is}}{\tilde{n}_1+\tilde{n}_{is}}.
\label{p}
\eea

This implies that the equation for the total density of objects
$\tn_{t} = \tn_1 + \tn_{is}$ is simply $d\tn_t/dp = \rho_0$.
This last quantity is nothing else than the survival probability
after $\ell$ steps for a random walk in a lattice with a density $n_t$
of traps.
In the initial regime $p \ll 1$ the spatial distribution of traps
is uncorrelated. It is well known that~\cite{Hughesbook}
for small trap density $n_t$,
\be
\rho_0=(1-n_t)^{\Ndis}.
\label{p00}
\ee
We remind that the number $\Ndis$ of distinct sites visited by a
walker during $\ell$ steps is by definition the capture area $c$.
Using Eq.~(\ref{p00}) and the condition $n_t \ll 1$,
Eqs.~(\ref{meanf_isl}) become the rate 
equations~(\ref{meanf}) valid for small $p$, as expected.
The total density obeys then
\be
\frac{d\tilde{n}_t}{dp}=\exp(-\tilde{n}_t)
\label{meanf_tot}
\ee
with solution
\be
\tilde{n}_t=\log(1+p).
\label{sol_n_tot}
\ee

\begin{figure}
\includegraphics[angle=0,width=13cm,clip]{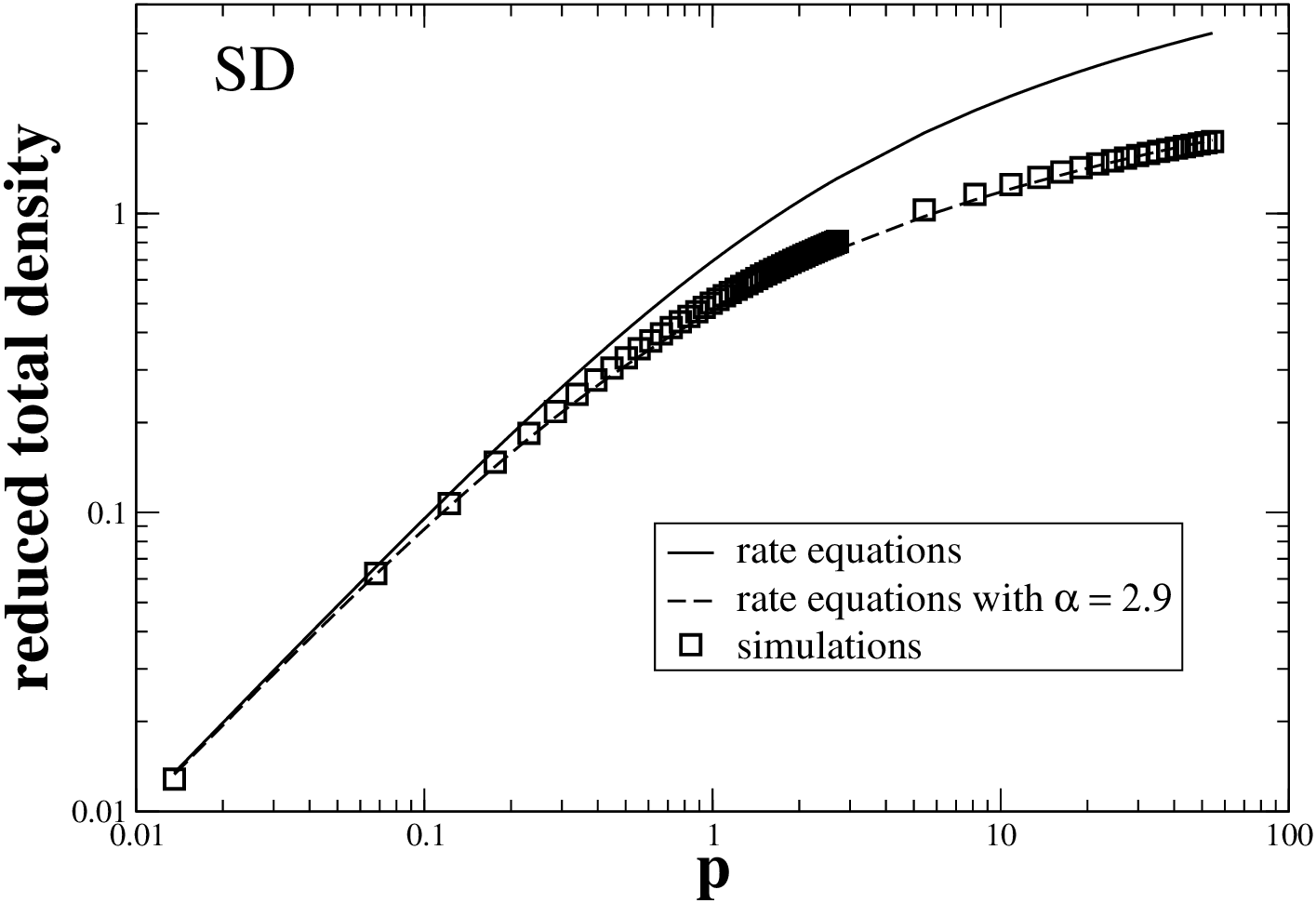}
\caption{
Behavior of $\tn_t$ for the SD model:
the square symbols are the results of numerical simulations for $\ell=100$;
the solid line is the theoretical prediction according to
Eq.~(\ref{sol_n_tot}), the dashed line given by
Eq.~(\ref{true_sol}) with $\alpha \approx 2.9$.}
\label{fig_objdens}
\end{figure}

Equation~(\ref{sol_n_tot}) is exact for $p \to 0$, but it makes
formally sense also for large $p$ where it predicts logarithmic growth,
in qualitative agreement with numerical results. However, a close
comparison shows a quantitative mismatch (Fig.~\ref{fig_objdens}).
The origin of this difference is clear: we have assumed an uncorrelated
distribution of objects. This is appropriate at the
beginning of the dynamics, but it becomes incorrect when the overlap
between capture areas starts to be significant.
To overcome this difficulty, let us study $G(x)$,
the distribution of distances
between nearest neighbor objects (both islands and adatoms).

Let us consider first the distance distribution $\Gunc(x)$ for objects put in
the lattice without correlations (with probability $n_t$).
In order for an object to be at distance $x$ from its nearest neighbour,
the $\nu(x)$ sites at distance $y<x$ must be empty: this happens
with probability $1-n_t$ for each of them. 
Moreover, at least one of the $\delta(x)$ sites at distance $x$ must be
occupied. 
Then
\be
\Gunc(x)\propto (1-n_t)^{\nu(x)}\cdot[1-(1-n_t)^{\delta(x)}]
\label{gauss0}
\ee

\begin{figure}
\includegraphics[angle=0,width=13cm,clip]{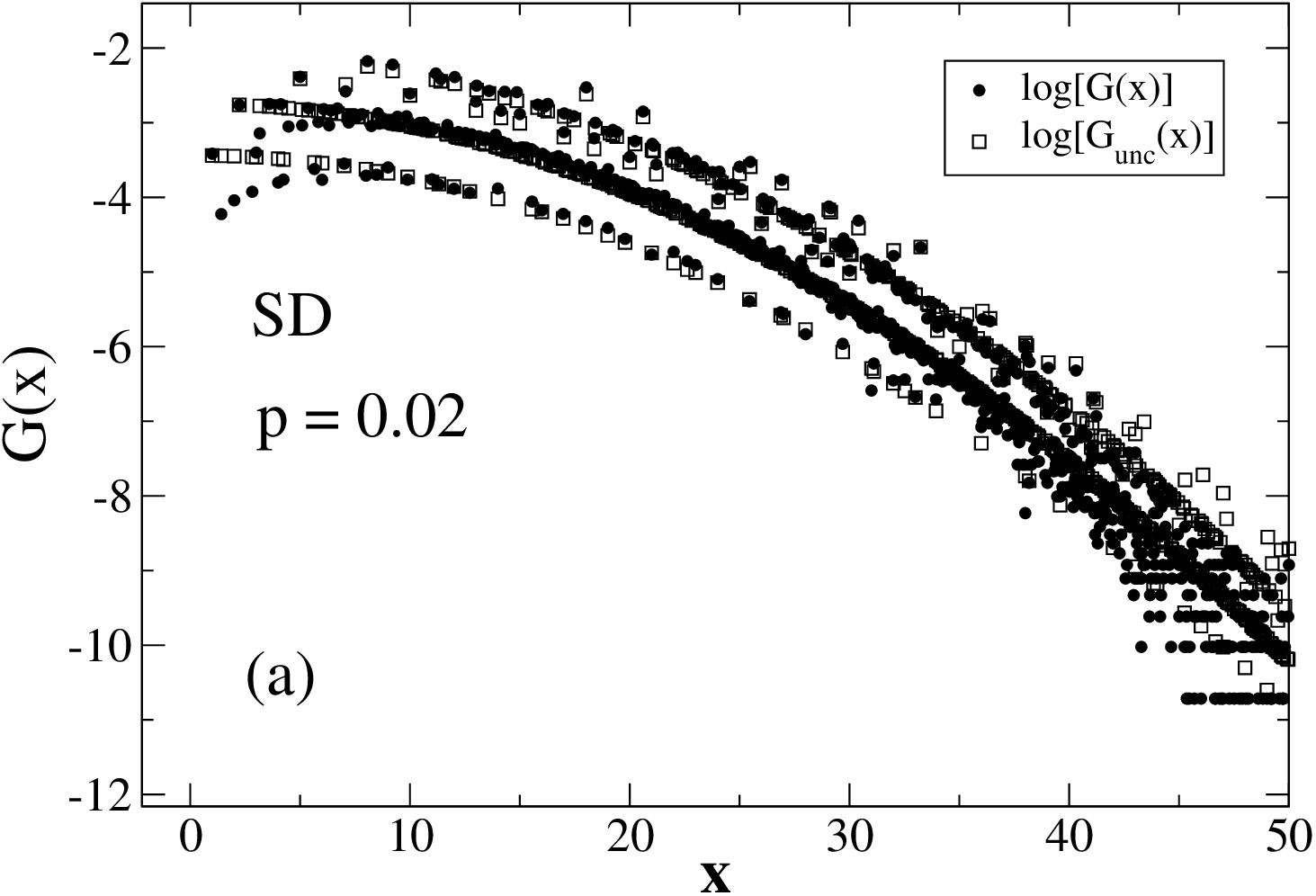}
\includegraphics[angle=0,width=13cm,clip]{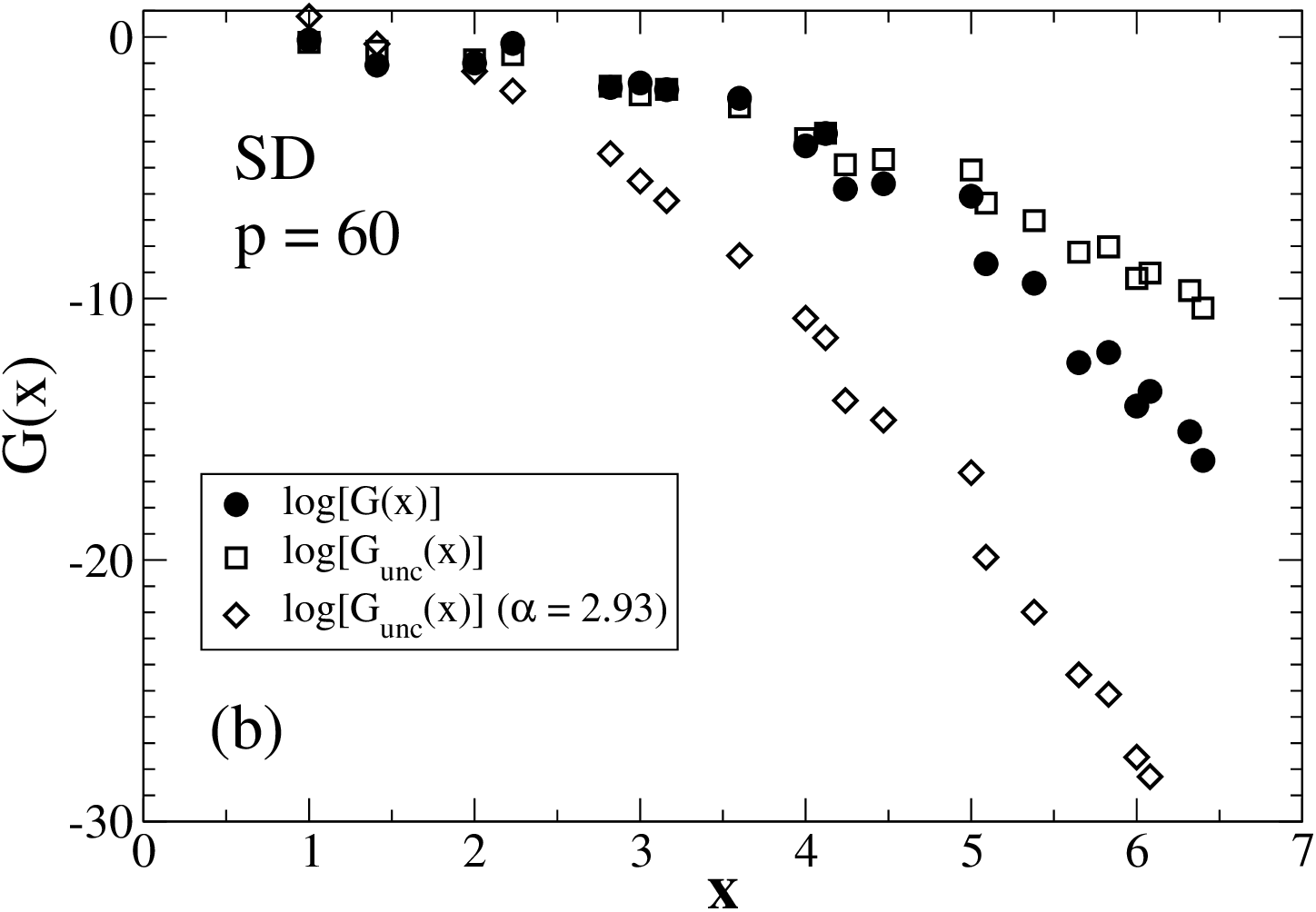}
\caption{(a) Comparison of $G(x)$ measured in simulations for the SD model
(filled circles) with the form predicted assuming lack of
correlations $\Gunc(x)$ 
[Eq.~(\ref{gauss0})] (empty squares) for $p=0.02$.
(b) Comparison of $G(x)$ measured in simulations for the SD model (filled
circles) with the form predicted assuming lack of correlations $\Gunc(x)$ 
[Eq.~(\ref{gauss0})] (empty squares) and $\Gunc(x)$ with
$\alpha \approx 2.9$ for $p=60$ (empty diamonds).}
\label{fig_Gx}
\end{figure}

The quantity $\nu(x)$ can be reasonably approximated with the continuum
expression $\nu(x)= \pi x^2$, resulting in a Gaussian contribution to
$\Gunc(x)$.
The other factor, equal to $2 \pi x$ in a continuous approximation,
strongly fluctuates for the different integers, due
to the discreteness of the lattice.
In Fig.~\ref{fig_Gx} we compare $G(x)$ with $\Gunc(x)$ for small
(a) and large (b) values of $p$.
While for $p \ll 1$ the agreeement is excellent~\footnote{The
agreement is very good but not perfect, even for $p \to 0$,
because for distances smaller than $\sqrt{c}$ correlations are always
at work and one expects a reduced probability $G(x)$ with respect to
the fully uncorrelated case.}, for
large $p$ there is a systematic deviation for large distances.

To understand which part of the distribution $G(x)$ of distances between
nearest neighbors is relevant for the determination of the
probability $\rho_0$ that enters Equations~(\ref{meanf_isl})
for the SD model in the late stage, let us consider first for simplicity
the ND model in one dimension.
In such a case, $\rho_0$ is the probability that a newly deposited particle
is at least at a distance $\ell$ from all islands in the system.
Contributions to $\rho_0$ are given only by the intervals between
particles larger than $2\ell$ so that
\be
\rho_0 = {1 \over L} \int_{2\ell}^{\infty} dx G(x) (x-2\ell)
\label{appeq}
\ee
For the SD model, due to the stochastic nature of diffusive motion
one cannot write down such a simple exact expression.
However, Eq.~(\ref{appeq}) remains approximately true provided
the capture area $2 \ell$ is replaced with the capture area for
the SD model $\Ndis(\ell)$.
In two dimensions additional complications arise, because of the angular
constraints.
Nevertheless it is intuitively clear that also in two dimensions
the chance for an adatom not to be incorporated depends on the existence
of regions that are sufficiently far from all particles, and hence
from the shape of the function $G(x)$ for large $x$.

In order to fit the large-$x$ decay of $G(x)$ with a form like
Eq.~(\ref{gauss0}) one must multiply the density by a factor
$\alpha(p)$ that is close to 3 for large $p$ (see Fig.~\ref{fig_Gx}).
Qualitatively it is possible to understand why $\alpha>1$.
The spatial distribution of islands obtained with the SD model
differs from the spatial distribution of the random deposition
model in one important respect: the probability to find very close islands
is strongly suppressed by the diffusion/aggregation
process, while it is significant in the random deposition model.
These neighboring islands do not increase very much the capture
probability, because they behave as a single trap. 
In simple words, for the same value of $p$, the islands of the
SD model are more effective to trap diffusing atoms than the islands
of the random deposition model: therefore, the latter model requires
a higher density of islands (i.e. a higher $p$) in order to get the
same trapping effect.

If now we take $\alpha$ to be constant, Eq.~(\ref{meanf_tot}) becomes
\be
\frac{d\tilde{n}_t}{dp}=\exp(-\alpha\tilde{n}_t)
\label{Ren_meanf_tot}
\ee

whose solution is
\be
\tilde{n}_t=\frac{1}{\alpha}\cdot\log(1+\alpha p).
\label{true_sol}
\ee

Such expression shows a much better agreement with
numerical results, as one can see in Fig.~\ref{fig_objdens}.
When $p$ becomes large enough, the density of adatoms is much smaller than
that of islands. Hence
\be
\tilde{n}_{is} \approx \tn_t =\frac{1}{\alpha}\log(1+\alpha p)
\label{high_p_is}
\ee
and, using Eq.~(\ref{meanf_isl}) and the fact that $\rho_{is} \simeq 1$,
\be
\frac{d\tilde{n}_{is}}{dp}=\rho_1\simeq\frac{\tilde{n}_1}{\tilde{n}_{is}}
\ee
from which we obtain
\be
\tilde{n}_1\simeq\tilde{n}_{is}\frac{d\tilde{n}_{is}}{dp}
\simeq\frac{1}{\alpha}\cdot\frac{\log(1+\alpha p)}{1+\alpha p}.
\label{high_p_ad}
\ee

In this way, we have recovered that, for $p \gg 1$,
$\tilde{n}_1\sim p^{-1}$ (apart from a logarithmic correction)
and $\tilde{n}_{is}$ is logarithmic in $p$ (see Fig.~\ref{dens}).

\section{Island Size Distribution}
\label{Sec5}

\begin{figure}
\includegraphics[angle=0,width=13cm,clip]{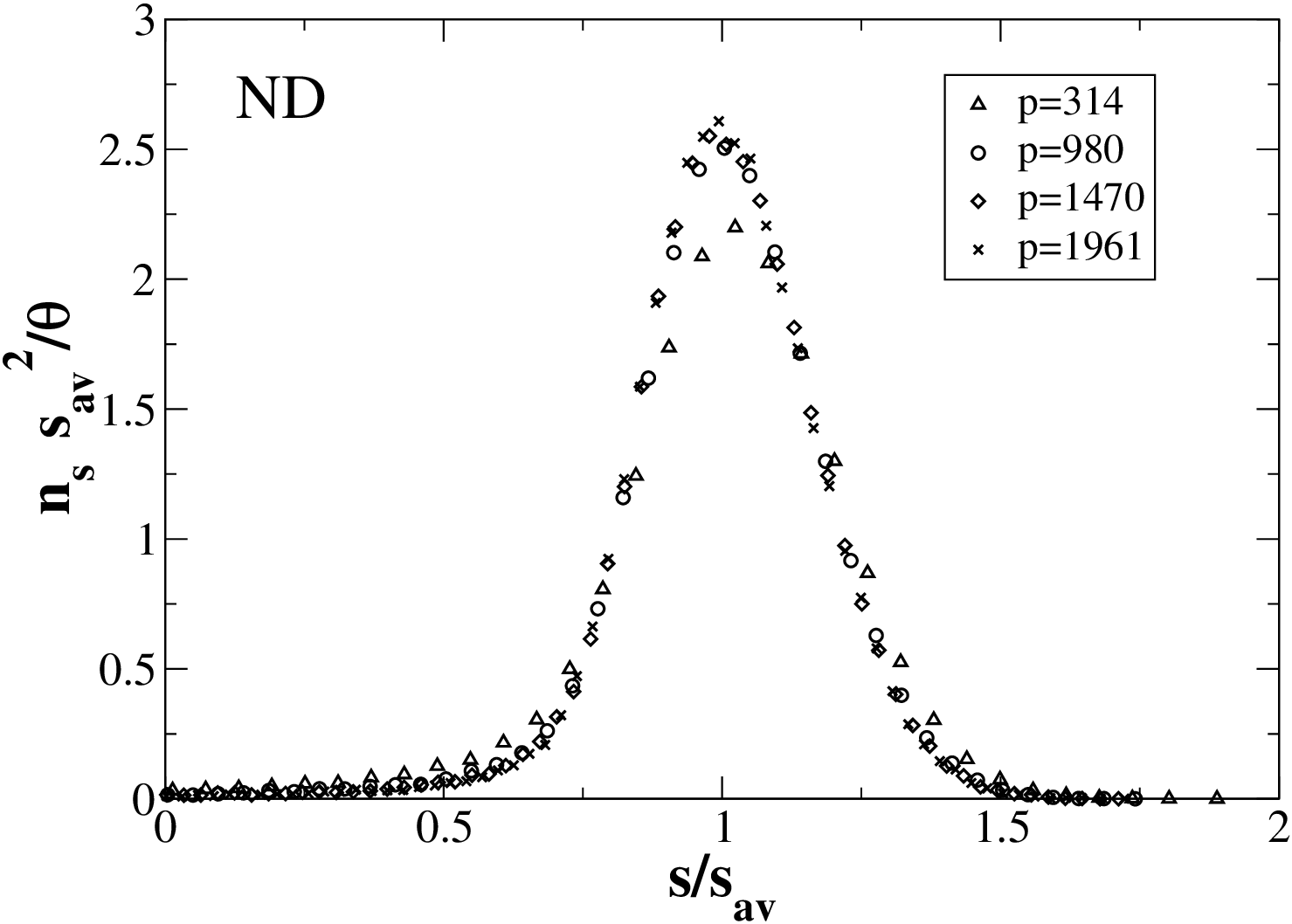}
\caption{Collapse plot of the Island size Distribution for the ND model.
Data are for $\ell=10$.}
\label{Scaling_ND}
\end{figure}

\begin{figure}
\includegraphics[angle=0,width=13cm,clip]{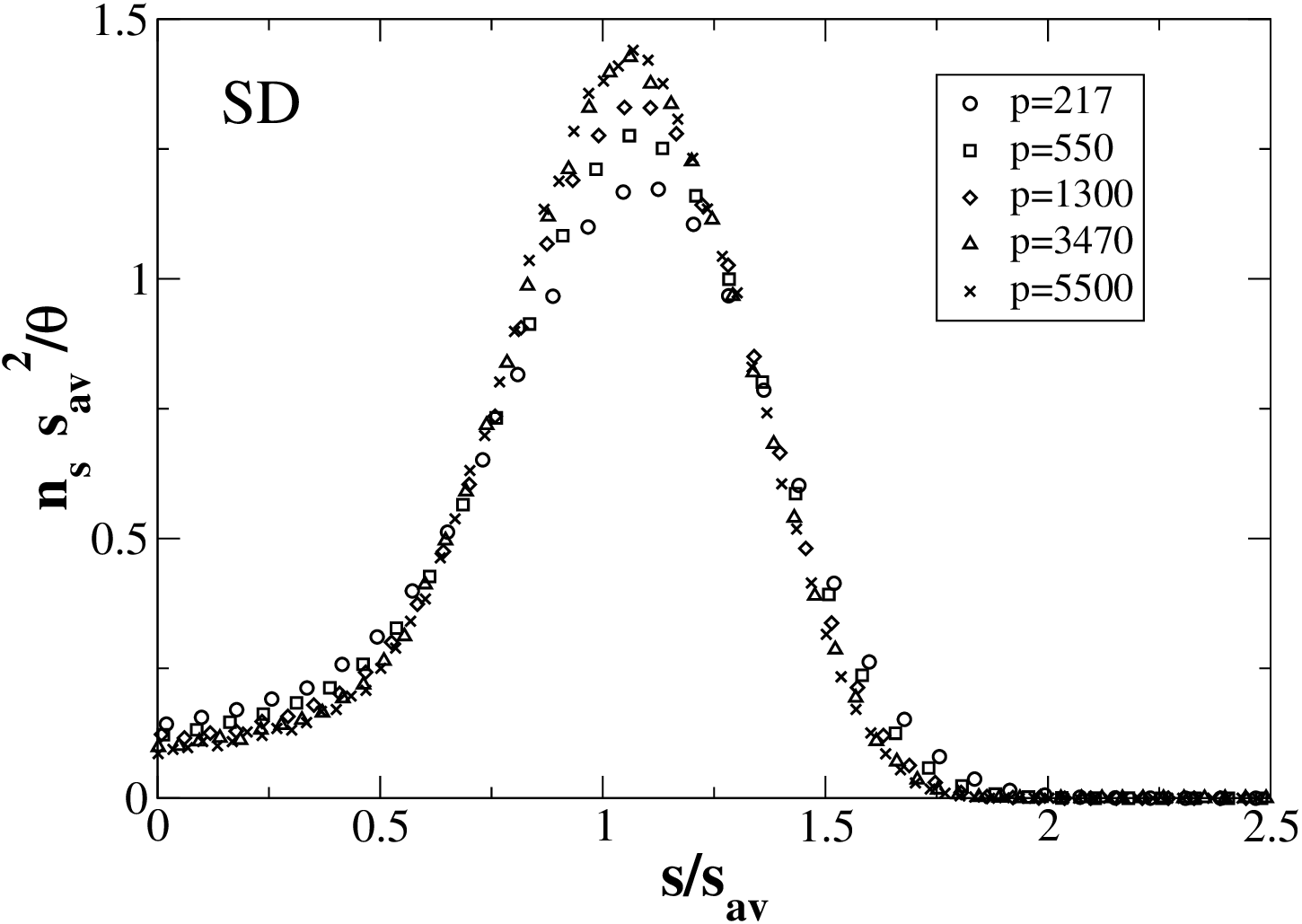}
\caption{Collapse plot of the Island size Distribution for the SD model.
Data are for $\ell=100$.}
\label{Scaling_SD}
\end{figure}

During the initial nucleation stage the island size distribution $n_s$
is given by Eq.~(\ref{n_s}), reflecting the lack of correlations in
the growth of different islands.
In the following steady-state regime the competition between different
islands strongly perturbs this form.
In the Full Diffusion model it is believed~\cite{Bartelt92} that the ISD
obeys asymptotically the scaling form
\be
n_s(\theta) = {\theta \over \sav^2(\theta)} f[s/\sav(\theta)]
\label{Eqscaling}
\ee
where
\be
\sav(\theta) = {\theta \over n_{is}(\theta)}
\label{s_av}
\ee
is the average size of an island.
In Figs.~\ref{Scaling_ND} and~\ref{Scaling_SD} we show that the same
scaling property holds for both models considered here, although
the approach to the asymptotic shape is much slower for the
Sequential Diffusion model.

While the scaling property is the same, the scaling function is rather
different in the two cases and this reflects the totally diverse
origin of scaling in the two cases.

For the No Diffusion model, scaling is a trivial consequence of the lack of
nucleation events in the steady state.
When the steady state is reached, the whole lattice is partitioned into
capture zones.
In the present case, owing to the deterministic nature of the
incorporation process, capture zones exactly coincide with Voronoi Cells.
During the steady-state, all deposited atoms are immediately
incorporated by the island owning the capture zone wherein the atom is
deposited.
Hence the growth rate of an island is (apart from fluctuations
in the deposition flux) perfectly proportional to the size of its capture zone.
If we wait long enough the island size distribution will then become equal
to the size distribution of capture zones.
In this sense scaling is trivial in the ND model. What is highly nontrivial
is the shape of the scaling function which is the outcome of the
complicated evolution during the nucleation stage~\cite{Mulheran96}.

As shown in Fig.~\ref{dens}, in the SD model island nucleation continues
during the steady-state, and if we wait long enough 
virtually all islands have been nucleated during such $p \gg 1$ regime.
It is clear that in this case the origin of the ISD scaling property
is different and has to do with the dynamics during the steady-state.
The situation is analogous to what happens in the Full Diffusion model.

\begin{figure}
\includegraphics[angle=0,width=13cm,clip]{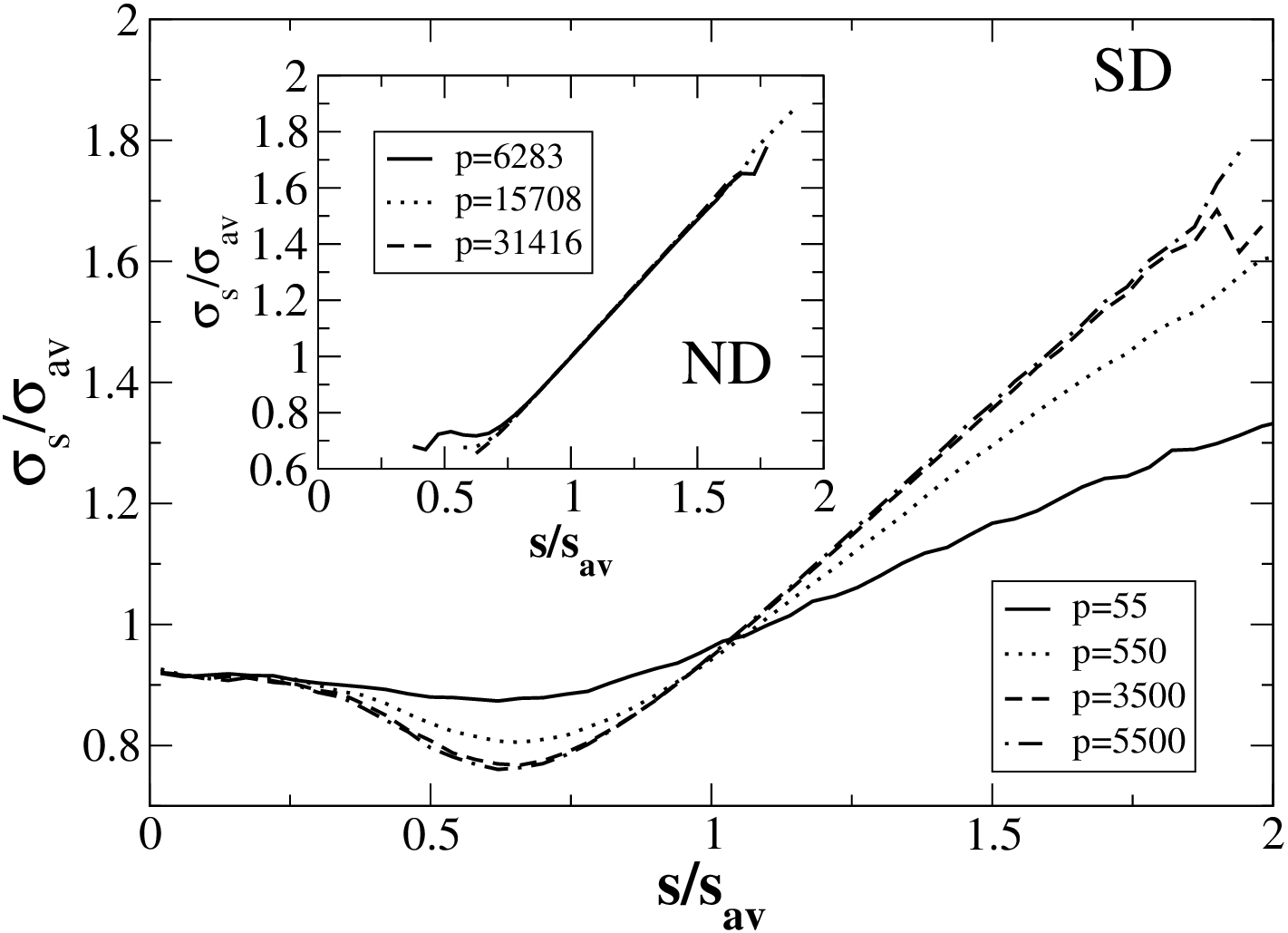}
\caption{Main: Scaling plot of the capture numbers according to
Eq.~(\ref{Scaling_form_C}) for the SD model.
Inset: Scaling plot of the capture numbers for the ND model.}
\label{Scaling_C}
\end{figure}

More insight into the problem is provided by the application of
the approach that Bartelt and Evans developed for Full Diffusion
models~\cite{Bartelt96}.
The basic idea is that scaling of the Island Size Distribution is
a consequence of a scaling relationship valid for the capture numbers
$\sigma_s$ entering the rate equations~(\ref{Rate_generic})
\be
{\sigma_s \over \sigmaav} = C\left({s \over \sav}\right).
\label{Scaling_form_C}
\ee
Introducing the scaling ansatz for $n_s$ and $\sigma_s$ into
the second of Eqs.~(\ref{Rate_generic}), and using the steady-state
condition, that in this case is $\sigmaav n_{is}=1$, a relation between the
scaling functions $f$ and $C$ is obtained~\cite{Bartelt96}
\be
f(x) = f(0) \exp\left[ \int_0^x dy {2 \omega-1-C'(y) \over C(y) - \omega y}
\right]
\label{Scaling_connection}
\ee
where 
\be
\omega = {d \ln \sav \over d \ln \theta}.
\ee
The analytical computation of the scaling form $C(x)$ for capture
numbers is a hard problem that has been approached
in many ways~\cite{Mulheran00,Evans01,Evans02,Mulheran04}.
Here we limit ourselves to its numerical evaluation, performed
in a way analogous to Ref.~\cite{Bartelt96}. Results are displayed in
Fig.~\ref{Scaling_C}.

It is clear that capture numbers for the SD differ much from the
mean-field assumption that they are all the same, i.e. independent of $s$.
More importantly, the scaling form~(\ref{Scaling_form_C})
tends to be fulfilled, but only in the limit of very large $p$.
This goes along with the observation that in the steady-state regime
$\sav$ grows linearly with $\theta$ but with a (sizeable) logarithmic
correction, so that $\omega$ in Eq.~(\ref{Scaling_connection})
tends to 1 with respect to $\theta$ only in the limit
$\theta \to \infty$. In particular, a linear fit for $s/\sav$ between 0.8
and 1.6 in Fig.~\ref{Scaling_C} yields the effective values
$\omega=0.40$ for $p=55$, $\omega=0.70$ for $p=550$,
$\omega=0.81$ for $p=3500$ and
$\omega=0.83$ for $p=5500$.
We conclude that deviations from scaling for the ISD displayed in
Eq.~(\ref{Scaling_SD}) are a consequence of the logarithmic corrections.
However the logarithmic growth of $n_{is}$, i.e. the fact that nucleations
continue during the steady-state, is also the very origin of the scaling
observed for SD. In the absence of nucleations one would have
$\omega=1$ and $C(x)=x$, as it happens for the ND model
(see the inset of Fig.~\ref{Scaling_C}).
If we insert such values in Eq.~(\ref{Scaling_connection}),
the integrand is indeterminate, confirming that the approach of
Ref.~\cite{Bartelt96} cannot be applied for ND, since the shape of the scaling
function $f(x)$ is not determined by the dynamics during the steady-state.

Finally, let us remark that the different nature of ISD scaling is also
reflected in the form of the scaling function $f(x)$. In the ND case,
since nucleation is inhibited there are no small islands, and
$f(s/\sav)$ vanishes for small argument.
Instead in the SD case, small islands are continuously created and
this implies that $f(0)$ is finite.

\section{Conclusions}

In this paper we have analyzed, by means of numerical simulations
and theoretical arguments, two models for irreversible submonolayer
growth that incorporate less fluctuations than the usual Full Diffusion model.
Let us conclude with some remarks on the findings.

As already mentioned in the Introduction, the definition of the
SD model seems similar to the model with critical size $i=0$,
also said spontaneous nucleation model.
Surprisingly, all results indicate that the two models
exhibit quite different behaviors.
In both models the trivial nucleation regime is followed by
a steady-state during which island nucleation continues, but in
the SD, island density grows logarithmically, while it grows as
a power law for SN~\cite{Blackman01}.
The difference is even larger for the Island Size Distribution.
At odds with the result~\cite{Amar95} that the scaling function for
spontaneous nucleation is a decreasing function of the
scaled size, i.e. its maximum is for $s/\sav \to 0$,
in the SD model the scaling function has a shape with a 
peak for finite $s/\sav$, much more similar to the Full Diffusion model.
This rather strong change can stem only from one of the two differences between
the models: the SD case is sequential (while the SN is not);
in the SD a particle diffuses for a fixed number of steps before
stopping, while this number fluctuates (with a well defined average) for SN.
The first difference is for sure irrelevant: also the SN model becomes
sequential in the limit $D/F \to \infty$ and its behavior does not change.
The key point is the second difference and in particular the fact that
in the SN there is high chance for an adatom to stop and nucleate
very early after its deposition.
This facilitates the nucleation of new islands also in the steady-state
and enhances the probability that newly formed islands have very
small capture zones and therefore relatively small sizes.
The fact that the walker must diffuse for $\ell$ steps turns out to
be crucial for having a peak for finite $s$ in the ISD
of the SD model.

Both point-island models analyzed here display scaling of the Island
Size Distribution in the steady-state.
This result is in apparent contradiction with the recent
statement~\cite{Ratsch05} that point-island models
cannot obey scaling, because of continuous nucleation
leading to a delta-like distribution of capture zone sizes.
According to this argument, scaling could occur only if nucleation of
new islands is suppressed
during the steady-state, so that each island collects a number of
adatoms proportional to its surrounding capture zone and the
scaling function is equal to the distribution of capture zone sizes.

The ND model obeys scaling, despite being a point-island model,
but this is not in real disagreement with Ref.~\cite{Ratsch05}.
The scaling observed is of the same type:
it simply reflects the proportionality between capture
zones and aggregation rates. It is associated with a linear
growth of the average island size $\sav$ as a function of coverage and
to a scaling function $f(x)$ that vanishes for $x \to 0$.
This may happen for point-islands because the particular
deterministic nature of capture areas in the ND model
fully inhibits nucleations in the steady state.

As shown in Fig.~\ref{Scaling_SD}, also the SD model obeys scaling.
This is a different type of scaling, that occurs just because
nucleation events occur during the steady-state.
This scaling is characterized by a sublinear growth of the
island density $\tn_{is}$ and by a finite value for $f(0)$ and
it is the same type of scaling that occurs for the FD model.

The origin of the two types of scaling is completely different:
while in the ND case the scaling function is determined
by what occurs during the nucleation stage, in the SD
any feature of the nucleation stage is completely erased
by the steady-state dynamics, the only thing that matters.
The second type of scaling is compatible with point-island models,
because the steady-state can be made infinitely longer than
the nucleation stage so that all memory of the previous regime is lost,
still keeping the coverage smaller than 1.
In the SD model this occurs considering $1/c(\ell) \ll \theta \ll 1$.
In the FD model this occurs when $(D/F)^{-1/2} \ll \theta \ll 1$.
It would be interesting to gain a greater insight into the SD model,
trying to determine analytically the form of the scaling function
$C$ and formally prove that scaling occurs.

\end{document}